\documentclass[conference]{IEEEtran}
\IEEEoverridecommandlockouts
\usepackage{cite}
\usepackage{amsmath,amssymb,amsfonts}
\usepackage{algorithmic}
\usepackage{graphicx}
\usepackage{textcomp}
\usepackage{xcolor}
 
\DeclareMathOperator*{\argmax}{argmax}

\def\BibTeX{{\rm B\kern-.05em{\sc i\kern-.025em b}\kern-.08em
    T\kern-.1667em\lower.7ex\hbox{E}\kern-.125emX}}

\usepackage[left=0.680in, right=0.680in, bottom=1.049in, top=0.7in]{geometry}    
    
\begin{document}

\title{Fair Resource Allocation in Optical Networks under Tidal Traffic
}
\author{\IEEEauthorblockN{Tania Panayiotou and Georgios Ellinas
\IEEEauthorblockA{KIOS Research and Innovation Center of Excellence \\
Department of Electrical and Computer Engineering, University of Cyprus\\
\{panayiotou.tania, gellinas\}@ucy.ac.cy}
}}

\maketitle

\begin{abstract}
We propose an $\alpha$-fair routing and spectrum allocation (RSA) framework for reconfigurable elastic optical networks under modeled tidal traffic, that is based on the maximization of the social welfare function parameterized by a scalar $\alpha$ (the inequality aversion parameter). The objective is to approximate an egalitarian spectrum allocation (SA) that maximizes the minimum possible SA over all connections contending for the network resources, shifting from the widely used utilitarian SA that merely maximizes the network efficiency. A set of existing metrics are examined (i.e., connection blocking, resource utilization, coefficient of variation (CV) of utilities), and a set of new measures are also introduced (i.e., improvement on connection over- (COP) and under-provisioning (CUP), CV of unserved traffic), allowing a network operator to derive and evaluate in advance a set of $\alpha$-fair RSA solutions and select the one that best fits the performance requirements of both the individual connections and the overall network. We show that an egalitarian SA better utilizes the network resources by significantly improving both COP (up to $20\%$) and CUP (up to $80\%$), compared to the utilitarian allocation, while attaining zero blocking. Importantly, the CVs of utilities and unserved traffic indicate that a SA that is fairest with respect to the amount of utilities allocated to the connections does not imply that the SA is also fairest with respect to the achievable QoS of the connections, while an egalitarian SA better approximates a fairest QoS-based SA.              
\end{abstract}

\begin{IEEEkeywords}
fairness, game theory analysis, optimization, routing and spectrum allocation, tidal traffic, optical networks 
\end{IEEEkeywords}

\section{Introduction}
Allocation of optical network resources has long been associated with the objective of maximizing the {\it network efficiency}, with the efficiency usually measured according to performance evaluation metrics reflecting the aggregated network throughput (e.g., connection blocking, spectrum utilization, etc.). Maximization of the network efficiency, however, lacks {\it fairness}, as it does not ensure that the available resources are fairly divided among the connections to achieve a similar quality-of-service (QoS) for all connections. Specifically, in a network environment, where connections are likely to contend for the same bandwidth resources, maximization of the network efficiency tends to favor connections with higher bit-rate tendencies, possibly blocking connections with lower bit-rate tendencies. Furthermore, the commonly adopted performance evaluation metrics do not account for the examination of fairness measurements, and hence optimization of such measurements is not taken into consideration during resource allocation.  

On this basis, in this work we examine fairness considerations according to an $\alpha${\it-fair} resource allocation scheme~\cite{Bertsimas12} that allows a number of fairness measures to be explicitly examined for a reconfigurable elastic optical network (EON) under tidal traffic. In general, tidal traffic refers to the repetitive spatio-temporal variations of traffic demand (i.e., fluctuations), caused mainly by the daily large-scale population migration among different areas at different times~\cite{Zhong16}, and enhanced by the emergence of new types of networks (e.g., 5G networks), applications (e.g., video streaming), and services (e.g., connected cars). As tidal traffic is present in optical transport segments, SDN-enabled reconfigurable optical networks capable of following as closely as possible tidal traffic, have recently gained attention from the research community~\cite{BYan18, Alvizu17, Morales17, Fernandes15, Ohsita07, Caballero14, Panayiotou18, Panayiotou19, Kokkinos19, Singh18}.

Specifically, related work deals with traffic demand prediction~\cite{BYan18,Alvizu17, Morales17,Fernandes15,Ohsita07,Panayiotou18, Panayiotou19,Singh18} and with predictions used as inputs to resource allocation problems~\cite{BYan18, Alvizu17, Morales17, Fernandes15, Ohsita07, Caballero14, Panayiotou18, Panayiotou19, Kokkinos19, Singh18}, allowing in advance network optimization. The performance evaluation metrics commonly considered are associated with the network throughput, that is measured in connection blocking percentage~\cite{Kokkinos19,BYan18,Singh18, Panayiotou18, Panayiotou19}, spectrum utilization (e.g., number of frequency slots utilized)~\cite{Kokkinos19,Singh18,Panayiotou18, Panayiotou19}, and/or the degree of defragmentation~\cite{Kokkinos19}. However, such performance evaluation metrics lack fairness information, especially as it concerns the QoS of contending connections. Moreover, the commonly considered optimization objectives do not account for the fairness of the allocations. Only a few works exist in the literature focusing on fair resource allocation strategies~\cite{Callegati13,SMa13,Chen17}. However, in these works, the only measure of fairness considered is the blocking probability.  

To this end, we propose the development of an $\alpha$-fair routing and spectrum allocation (RSA) framework for reconfigurable EONs, under modeled tidal traffic, aiming to derive and evaluate in advance fair SAs. In this framework, an SDN-based central decision management unit (i.e., network orchestrator) decides on the allocation by maximizing the constant elasticity social welfare function $W_{\alpha}(u)$, parameterized by $\alpha \geq 0$ and defined for $u \in \mathbb{R}^n_+$, where $u$ is an $\alpha$-fair {\it utility allocation} for the $n$ players. In this work, the connections are perceived as the players competing for the link utilities, and the utilities are the frequency slots (FSs) available on the network links. The $\alpha$-fair RSA is formulated as an Integer Linear Program (ILP) that can be executed in advance, for each (re)configuration time interval of interest, allowing the reconfiguration to take place at the beginning of each time interval. 

Even though $\alpha$ is considered to be a natural measure of fairness~\cite{Bertsimas12}, in this work the $\alpha$-fair RSA is additionally evaluated according to several other, easier to interpret, measures, that also facilitate an understanding of the efficiency-fairness trade-offs for different $\alpha$ values. These measures include connection blocking percentage, resource utilization, and the CV of the utilities~\cite{Jain84}, as well as other measures that are introduced in this work aiming to directly interpret the fairness of the spectrum allocations (SAs) with respect to the QoS of contending connections. In particular, we introduce the improvement on connection over- (ICOP) and under-provisioning (ICUP), and the CV of the expected unserved traffic. These measures can be evaluated in advance, for several $\alpha$ values, according to the fluctuations resulting from the modeled tidal traffic. This allows the centralized controller to decide the $\alpha$-fair RSA solution that best meets the performance requirements of both the connections and the network, before network (re)configuration takes place.   
      
Overall, we show that as $\alpha$ increases, the proposed ILP-based $\alpha$-fair RSA algorithm approximates an egalitarian SA (i.e., an allocation that maximizes the minimum possible SA over all connections contending for the network resources) by significantly improving connection blocking, COP and CUP, compared to the widely used utilitarian allocation (i.e., an allocation that maximizes the network efficiency). Furthermore, the CV of utilities indicates that a SA that is fairest with respect to the amount of allocated utilities, does not imply that the SA is also fairest with respect to the QoS that contending connections enjoy. Importantly, the CV of unserved traffic indicates that an egalitarian allocation best approximates a SA that is fairest with respect to the connections' QoS. It is worth mentioning that $\alpha$-fair schemes have been previously examined in wireless networks~\cite{Altman08}, SDN IP networks~\cite{Allybokus18}, and TCP networks~\cite{Lan10, Tang06,Buzna17}. In optical networks, however, $\alpha$-fairness has not been previously considered.

\section{${\bf \alpha}$-Fairness Preliminaries}\label{prel} 
According to the $\alpha$-fair scheme, a central decision management unit decides on the allocation of the available resources among $n$ players by maximizing the constant elasticity welfare function $W_{\alpha}$ parameterized by $\alpha \geq 0$~\cite{Bertsimas12} as

\begin{equation}
\small
\quad W_{\alpha}(u) = \left\{
	\begin{array}{ll}
		\sum\limits_{i=1}^n \frac{u_i^{1-\alpha}}{1-\alpha}  & \mbox{for} \quad \alpha \geq 0, \alpha \neq 1 \\\\
		\sum\limits_{i=1}^n \log(u_i) & \mbox{for} \quad \alpha = 1
	\end{array}
\right.
\label{eq_alpha}
\end{equation}

\noindent where $u \in \mathbb{R}_+^n$ is a {\it utility allocation} and $u_i$ is equal to the utility derived by player $i$. In general, a utility allocation is {\it feasible} if and only if there exists an allocation of resources for which the utilities derived by the players are $u_1,u_2,...,u_n$, accordingly. The utility set $\mathcal{U} \subset \mathbb{R}_+^n$ is then defined as the set of all feasible utility allocations, and a resulting {\it $\alpha$-fair allocation}, denoted by $\pi(\alpha)$, is such that $\pi(\alpha) \in \argmax_{u \in \mathcal{U}} W_{\alpha}(u)$.

An explanation of why $W_{\alpha}$ yields fair allocations is that it exhibits diminishing marginal welfare increase, as utilities increase (because $W_{\alpha} $ is concave and component-wise increasing), with parameter $\alpha$ controlling the rate at which marginal increases diminish~\cite{Bertsimas12}. For example, if player $i$ is allocated less utilities than player $j$, then a marginal increase in the utilities of player $i$ (as $\alpha$ increases) would yield a higher welfare increase compared to a marginal increase in the utilities of player $j$. Hence, the marginal increase in the utilities of player $i$ would be more preferable. 

Special cases of the $\alpha$-fairness scheme are derived for $\alpha=0$, $\alpha=1$, and $\alpha \rightarrow \infty$. Specifically, for $\alpha=0$ the $W_{\alpha}$ function corresponds to finding a {\it utilitarian allocation} of the associated optimization problem; that is, an optimal solution of the problem that results in the highest system efficiency (e.g., maximum network throughput). For $\alpha=1$ the scheme corresponds to {\it proportional fairness} (the generalization of the Nash solution for a two-player problem)~\cite{Bertsimas11}, and for $\alpha \rightarrow \infty$ the scheme converges to {\it max-min fairness}; that is, an egalitarian allocation that ensures that the minimum allocated utility is as high as possible. Hence, a higher $\alpha$ value corresponds to a fairer scheme~\cite{Bertsimas12}. It is worth noting that utilitarian, proportional, and egalitarian schemes yield Pareto optimal allocations~\cite{Bertsimas11}. For our network application scenario, Pareto optimality means that the connections that do not contend with other connections along their routes (links) will be allocated their maximum requested rates, subject to the link capacity constraints.     

\section{Problem Statement}\label{PS} 
Given an EON with $k$ links and $M$ FSs at each network link, the general objective is to find an $\alpha$-fair RSA for $n$ connection requests. For solving the conventional RSA problem, a feasible route and a SA must be found for each connection request, forming a lightpath. For the proposed $\alpha$-fair RSA the assumption is that the routes are pre-computed and given in the link utilization matrix $P=[p_{il}] \in \mathbb{R}^{n \times k}$, indicating the contending connections for each link in the network (i.e., the connections utilizing the same links). Therefore, in this work, the utilities are the available FSs, and the players are the connections. The utilities are given in the SA matrix $U=[u_{ij}] \in \mathbb{R}_+^{n \times m}$, consisting of all the possible SAs (actions) for each connection $i$. Hence, $u_{ij}$ is SA $j$ for connection $i$, measured in the number of FSs, and $m$ is the number of possible SAs for each connection $i$. Since $M$ is the available link bandwidth, then $0 \leq u_{ij} \leq M$, for all connections and SAs.

Given matrices $P$ and $U$, the {\it objective} is to obtain an egalitarian $\alpha$-fair SA subject to the feasibility constraints of the EON technology. The problem is formulated as an ILP with the objective of maximizing the social elasticity welfare function, $W_{\alpha}(\hat{u})$, parameterized by $\alpha \geq 0$ and defined for $\hat{u} \in \mathbb{R}_+^n$, where $\hat{u}=[\hat{u}_1,..,\hat{u}_n] \in \mathcal{\hat{U}}$ is a feasible SA of the normalized utilities $\hat{u_i} \in \hat{U}$, derived by normalizing the SA matrix $U$. Note that, by normalizing the utilities we capture the fact that we are less concerned with the amount of utility achieved by a connection and more with the fraction of maximum possible utility that a connection achieves (i.e., the satisfaction of an end-user); a common practice in welfare maximization problems that focus on approximating egalitarian welfare~\cite{Aziz19}. Hence, an $\alpha$-fair allocation of the normalized utilities is given by $\hat{\pi}(\alpha) \in \argmax_{\hat{u} \in \mathcal{\hat{U}}} W_{\alpha}(\hat{u})$, from which the $\alpha$-fair allocation of the utilities, $\pi(\alpha)$, can be derived. Note that sets $\mathcal{U}$ and $\mathcal{\hat{U}}$ are derived during the execution of the ILP algorithm. 

\begin{figure}[ht]
\begin{center}
\includegraphics[width=\columnwidth]{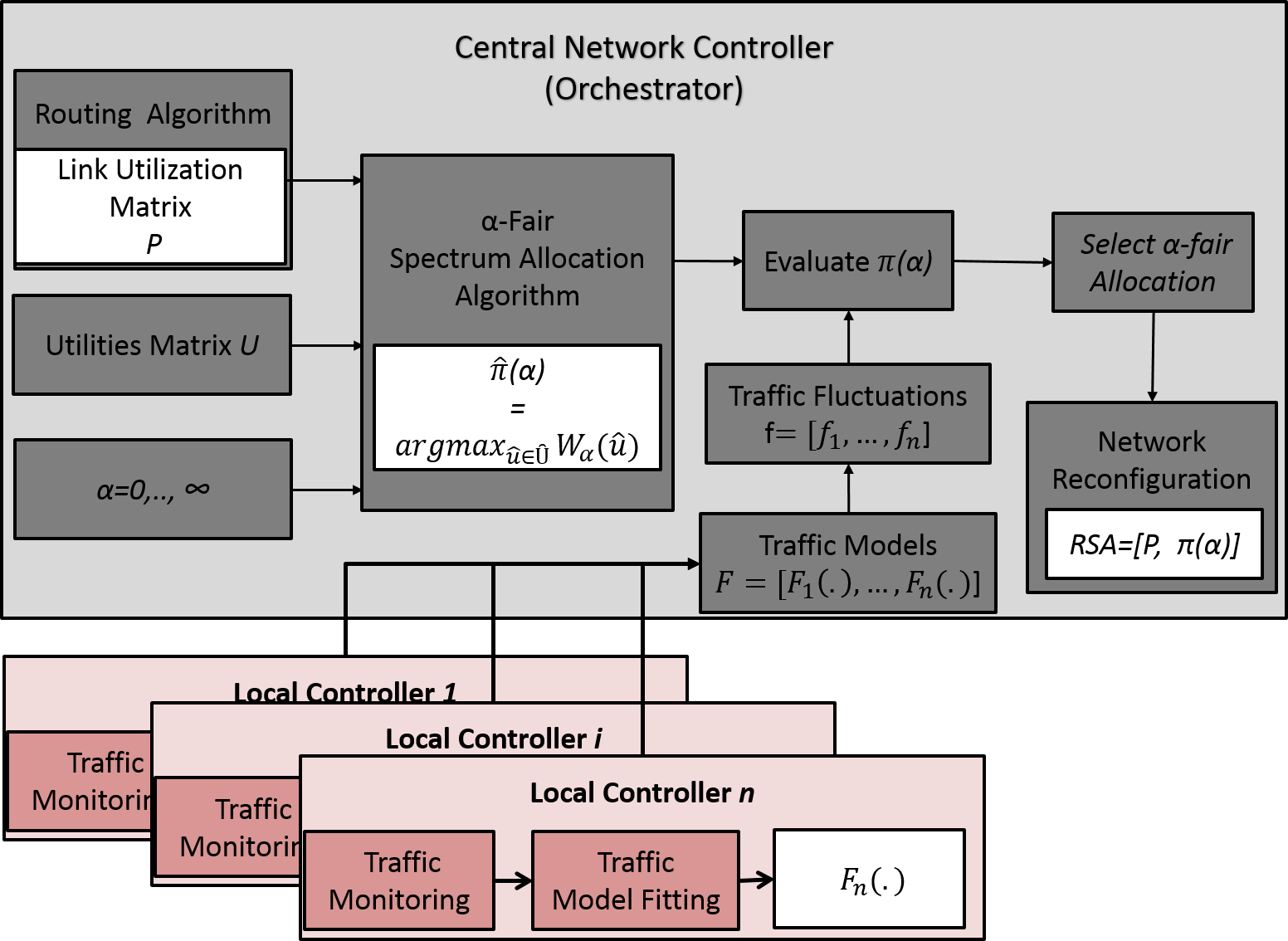}
\caption{$\alpha$-fair RSA framework.}
\label{framework}
\end{center}
\end{figure} 

Figure~\ref{framework} illustrates the proposed $\alpha$-fair RSA framework. This framework is performed off-line, for each time interval of interest (e.g., hourly, daily time intervals), and the network configurations are readily available at the beginning of each time interval. For simplicity, time interval notations are omitted from Fig.~\ref{framework}. The $\alpha$-fair RSA algorithm is executed for several $\alpha$ values, and each $\alpha$-fair allocation $\pi(\alpha)$ is evaluated in advance. Hence, the network operator is capable of selecting the $\alpha$-fair RSA that best meets specific performance targets.  

Specifically, $\alpha$-fair allocations are evaluated according to the connection blocking percentage, resource  utilization, CV of utilities, CV of expected unserved traffic, and the improvement on COP/CUP. The latter measures are based on the expected unserved traffic and excess utilities, $u^-=[u_1^-,..,u_n^-]$ and $u^+=[u_1^+,..,u_n^+]$, respectively, estimated according to the a-priori modeled tidal traffic. Tidal traffic, is given in a set of traffic demand distributions $F=[F_1(\cdot),..,F_n(\cdot)]$, from which traffic fluctuations $f=[f_1,...,f_n]$ are drawn as $f\sim F$. Hence, $u^-$ and $u^+$ are estimated by the fluctuations around $u \in \pi(\alpha)$. Note that each distribution is assumed to be a function of the bandwidth demand measured according to the requested FSs. Even though the requested number of FSs depends on several factors, such as the modulation format, in this work, similar to~\cite{Panayiotou19}, for simplicity we assume that the distributions directly reflect the requested number of FSs (i.e., logical link usage); an assumption that does not affect the scope of this work.  

According to Fig.~\ref{framework}, traffic demand models, $F$, are present in local controllers (i.e., offloading the central controller from functionalities that can be performed locally) by means of monitoring and analyzing the aggregated traffic of each source-destination pair forming a logical connection. After model fitting, the relevant parameters (e.g., distribution, mean, and standard deviation) are communicated to the central controller, from which fluctuations $f$ can be drawn. The reader should note here that the traffic modeling procedure is omitted as it is out of the scope of this work.

\section{$\alpha$-Fair Routing and Spectrum Allocation}
The $\alpha$-fair RSA algorithm is based on the maximization of $W_{\alpha}(\hat{u})$, where $\hat{u}=[\hat{u}_1,..,\hat{u}_n]\in \mathcal{U}$, and $\hat{u}_i \in \hat{U}$ are the normalized utilities of SA matrix $U$. Specifically,\\
$\bullet$ $U=[u_{ij}] \in  \mathbb{R}^{n\times m}$ is the matrix describing all the possible SA $j$ of each connection $i$, for a total of $m$ possible SAs, and

\begin{equation}
\small
 u_{ij}=\left\{
	\begin{array}{ll}
		j\times \frac{M}{m}   & \mbox{if } (j\times \frac{M}{m}) \leq u_{imax} \\\\
		0 & \mbox{otherwise}
	\end{array}
\right.
\label{eq0}
\end{equation}

\noindent where $u_{imax}$ is the peak-rate demand of connection $i$ (i.e., given by $F_i(\cdot)$). According to Eq.~(\ref{eq0}), the number of FSs increases as index $j$ increases, with a maximum possible allocation equal to the maximum link capacity, $M$. For each connection $i$, $u_{ij}$ is set to $0$ when $j\times M/m$ is greater than its peak-rate demand. By doing so, we avoid including in $U$ allocations that will lead, with certainty, to unnecessary connection overprovisioning. \\   
$\bullet$ $\hat{U}=[\hat{u}_{ij}] \in  \mathbb{R}^{n\times m}$ is the SA matrix of the normalized utilities, where

\begin{equation}
\small
 \hat{u}_{ij}=\left\{
	\begin{array}{ll}
		\frac{u_{ij}}{u_{imax}}   & \mbox{if } u_{ij} > 0 \\\\
		\epsilon & \mbox{otherwise}
	\end{array}
\right.
\label{eq1}
\end{equation}

\noindent and $\epsilon$ is a small positive value used for ensuring that $\hat{u} \in \mathbb{R}_+^n$. Note that $\epsilon$ must be greater than $0$ and less than the smallest possible SA (i.e., $\epsilon << M/m$). The denominator $u_{imax}$ is used for normalizing the utilities, such that $W_{\alpha}$ takes into account the normalized utilities $\hat{u} \in [\epsilon,1]$.  

For the ILP formulation, we assume that the routing problem is solved using Dijkstra's algorithm~\cite{Dij}, and \\
$\bullet$ $P=[p_{il}] \in \{0,1\}^{n\times k}, \hspace{0.05in} i=1,..,n, \hspace{0.05in} l=1,..,k$ is the link utilization matrix formed according to the pre-computed routes; $p_{il}=1$ if connection $i$ utilizes link $l$, and $0$ otherwise.\\
The variables, objective, and constraints of the ILP are defined as follows:\\
\\
\textbf{Variables}:\\
$\bullet$ $x_{ij}$: Boolean variable equal to $1$ if SA $j$ is chosen for connection $i$, and $0$ otherwise.\\
$\bullet$  $y_{is}$: Boolean variable equal to $1$ if FS $s$ is utilized by connection $i$, and $0$ otherwise.\\ 
$\bullet$  $z_{is}$: Boolean variable equal to $1$ if FS $s$ is the first FS utilized by connection $i$ amongst a set of contiguous FSs allocated to connection $i$, and $0$ otherwise.\\
\\
\\
\\
\textbf{Objective}:
\vspace{-0.1in}
\begin{equation}
\small
\textit{Maximize:} \quad W_{\alpha}(\hat{u}) = \left\{
	\begin{array}{ll}
		\sum\limits_{ij} x_{ij} \frac{\hat{u}_{ij}^{1-\alpha}}{1-\alpha}  & \mbox{if } \alpha \geq 0, \alpha \neq 1 \\\\
		\sum\limits_{ij} x_{ij}\log(\hat{u}_{ij}) & \mbox{if } \alpha = 1
	\end{array}
\right.
\label{obj}
\end{equation}
\textit{Subject to}:

\begin{equation}
\small
\sum\limits_{ij} u_{ij} x_{ij} p_{il} \leq M, \quad \forall \hspace{0.05in} l=1,..,k  
\label{con_1}
\end{equation}
\begin{equation}
\small
\sum\limits_{j} x_{ij} \leq 1,  \quad \forall \hspace{0.05in} i=1,...,n 
 \label{con_2}
\end{equation}
\begin{equation}
\small
\sum\limits_{s} y_{is} = \sum\limits_{j} u_{ij} x_{ij}, \quad \forall \hspace{0.05in} i=1,..,n  \label{con_3}
\end{equation}
\begin{equation}
\small
\sum\limits_{i} y_{is} p_{il} \leq 1, \quad \forall \hspace{0.05in} s=1,..,M, \hspace{0.05in} l=1,..,k   \label{con_4}
\end{equation}
\begin{equation}
\small
\sum\limits_{s} z_{is} \leq 1, \quad \forall \hspace{0.05in} i=1,...,n  
\label{con_5}
\end{equation}
\begin{equation}
\small
y_{is}-y_{i(s-1)} \leq z_{is}, \quad \forall \hspace{0.05in} i=1,..,n, \hspace{0.05in}  s=2,...,M  
\label{con_6}
\end{equation}
\begin{equation}
\small
y_{is}=0 \quad \text{for} \quad s=1, \quad \forall \hspace{0.05in} i=1,...,n
\label{con_7}
\end{equation}

The objective of the ILP is to maximize $W_{\alpha}(\hat{u})$, resulting in an $\alpha$-fair allocation $\hat{\pi}(\alpha)= \argmax_{\hat{u} \in \mathcal{\hat{U}}} W_{\alpha} (\hat{\hat{u}})$ of the normalized utilities, from which the actual $\alpha$-fair SA $\pi(\alpha)=u \in \mathcal{U}$ can be derived. Constraint~(\ref{con_1}) ensures that the SA decisions meet the link capacity constraints, while constraint~(\ref{con_2}) ensures that at most one allocation is chosen for each connection. Constraint~(\ref{con_3}) ensures that the number of allocated FSs is equal to the number of FSs indicated by the SA implemented and constraint~(\ref{con_4}) ensures that the no-frequency overlapping constraint is met for all network links. Finally, constraints~(\ref{con_5}),~(\ref{con_6}), and~(\ref{con_7}) ensure that each connection is assigned contiguous FSs on all links of its route.

\section{Evaluation Measures}
Once the $\alpha$-fair RSA is derived, the expected unserved traffic and excess utilities $u^-=[u_1^-,..,u_n^-]$ and $u^+=[u_1^+,..,u_n^+]$, respectively, are computed as 

\begin{equation}
\small
u^+_{i}= \frac {1}{T} \sum\limits_{t \mid u_{i} > f_{it} } (u_{i} - f_{it})
\end{equation}

\begin{equation}
\small
u^-_{i}=\frac{1}{T} \sum\limits_{t \mid u_{i} < f_{it} } \mid (u_{i} - f_{it}) \mid
\end{equation}

\noindent where $u_i \in \pi(\alpha)$ and $f_{it} \sim F_i(\cdot), \hspace{0.05in} \forall \hspace{0.05in} t=1,..,T$. Hence, $u^+$ and $u^-$ are evaluated according to $T$ traffic demand fluctuation around the $\alpha$-fair utilities $u \in \pi(\alpha)$. 

COP and CUP are introduced as measures of the aggregated expected excess utilities and unserved traffic, respectively, and are given by $COP(\alpha)=\sum_{i} u^+_i$ and $CUP(\alpha)=\sum_{i} u^-_i$. Improvements on COP (ICOP) and CUP (ICUP) are given by 

\begin{equation}
\scriptsize
ICOP(\alpha)=\frac{COP(0)-COP(\alpha)}{COP(0)}, \quad ICUP(\alpha)=\frac{CUP(0)-CUP(\alpha)}{CUP(0)}
\end{equation}

\noindent where $COP(0)$ and $CUP(0)$ are derived by the utilitarian allocation $\pi(0)$, which is likely to cause the highest values of COP and CUP. 
As $\alpha$ increases (a fairer allocation is derived), $ICOP(\alpha)$ is likely to only positively increase (up to one), as the contesting connections with higher bit-rate tendencies (over-provisioned) are likely to be allocated less resources, hence improving the overall COP. On the other hand, as $\alpha$ increases, $ICUP(\alpha)$ is likely to positively increase (as more connections are admitted into the network), without any guarantees however that it will not start to decrease as $\alpha \rightarrow \infty$ (e.g., max-min fairness is achieved by increasing under-provisioning). The latter depends on the network load and the traffic demand behavior of each connection. Nevertheless, the evaluation of these measures allows the network operator to decide on the allocation that best meets specific performance criteria for the connections and the overall network.  

Furthermore, even though parameter $\alpha$ is considered to be a natural measure of fairness, in this work we also use the CV of the utilities~\cite{Jain84}, to directly interpret the fairness of the allocations with respect to the amount of utilities (FSs) allocated to the connections. In particular, the CV of the utilities measures the dispersion of utilities allocated to the connections around the mean of the allocated utilities and is given by 

\begin{equation}
\small
CV(\alpha)=\sqrt{\frac{1}{n-1}\sum\limits_{i=1}^n \frac{(u_i-\bar{u})^2}{\bar{ u^2}}}
\label{CV}
\end{equation}

\noindent where $u_i \in \pi(\alpha)$, and $\bar{u}= \frac{1}{n} \sum_{i=1}^n \pi_i(\alpha)$ is the mean of allocated utilities. An $\alpha$-fair allocation is said to be fairer with respect to the utilities, if and only if $CV(\alpha)$ is smaller than the CVs of other $\alpha$ values. Hence, as $\alpha$ increases, CV($\alpha$) is likely to decrease. As we are also interested in investigating the fairness of the allocations with respect to the connections' achievable QoS, we also use the CV of the unserved traffic, evaluated according to Eq.~(\ref{CV}) as a function of $u^-$, instead of $u$. 
The $\alpha$-fair allocation that minimizes the CV of $u^-$ is the one
that best approximates the fairest allocation with respect to the connections' QoS. Note that it is more critical to derive a fair $u^-$, instead of a fair $u^+$, as connection under-provisioning is a noticeable and annoying effect for the end-users, while connection over-provisioning just tends to waste resources. Finally, regarding connection blocking, a connection is considered blocked when allocated $0$ FSs (i.e, $u_i=0$), and resource utilization is measured as the sum of the FSs (utilities) allocated to all connections along their routes.

\section{Performance Evaluation}\label{PE}
\begin{figure}[ht]
\begin{center}
\includegraphics[width=0.45\columnwidth]{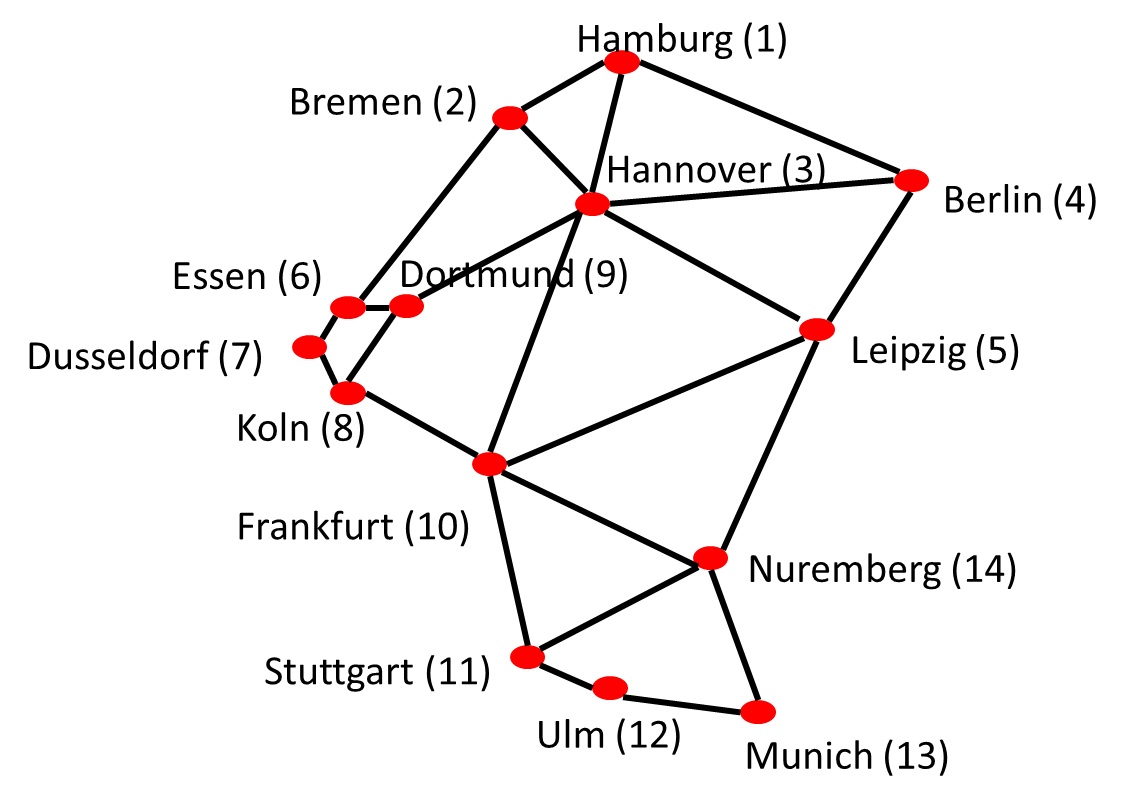}
\caption{Deutsche Telekom network topology.}
\label{net}
\end{center}
\end{figure}

In order to evaluate the performance of the proposed framework, the generic Deutsche Telekom (DT) network (14 nodes, 23 links) of Fig.~\ref{net} was used, with each FS set at $12.5$GHz and with each fiber link utilizing $M=100$ FSs. In total $m=50$ possible SAs were assumed for each one of the $n=20$ connections, with randomly selected source-destination pairs. 

For the evaluation of $u^+$ and $u^-$, $T=1000$ fluctuations ($f$) were drawn from a set of log-normal distributions $F$. Specifically, for each connection $i$, $f_{i} \sim F(\mu_{i},\sigma_{i}^2)$, with each $f_{i} \in (0,M)$. Parameters $\sigma_{i}^2$ and $\mu_i$, were drawn from uniform distributions $unif(0,1)$ and $unif(2.5,4.5)$, respectively. For the simulations, fluctuations $f_i$ were scaled down by a factor of two in order to fit the link capacity constraints. Further, the scaled fluctuations indicating a higher spectrum demand than $M$ were set to $f_i=M$. It is important to note that the log-normal distribution is used, as it has been shown that Internet traffic volumes are best characterized by such a distribution~\cite{Antoniou02, Kassim15} (i.e., by skewness and heavy tails).

\begin{figure}[h!]
\begin{center}
\includegraphics[width=0.91\columnwidth]{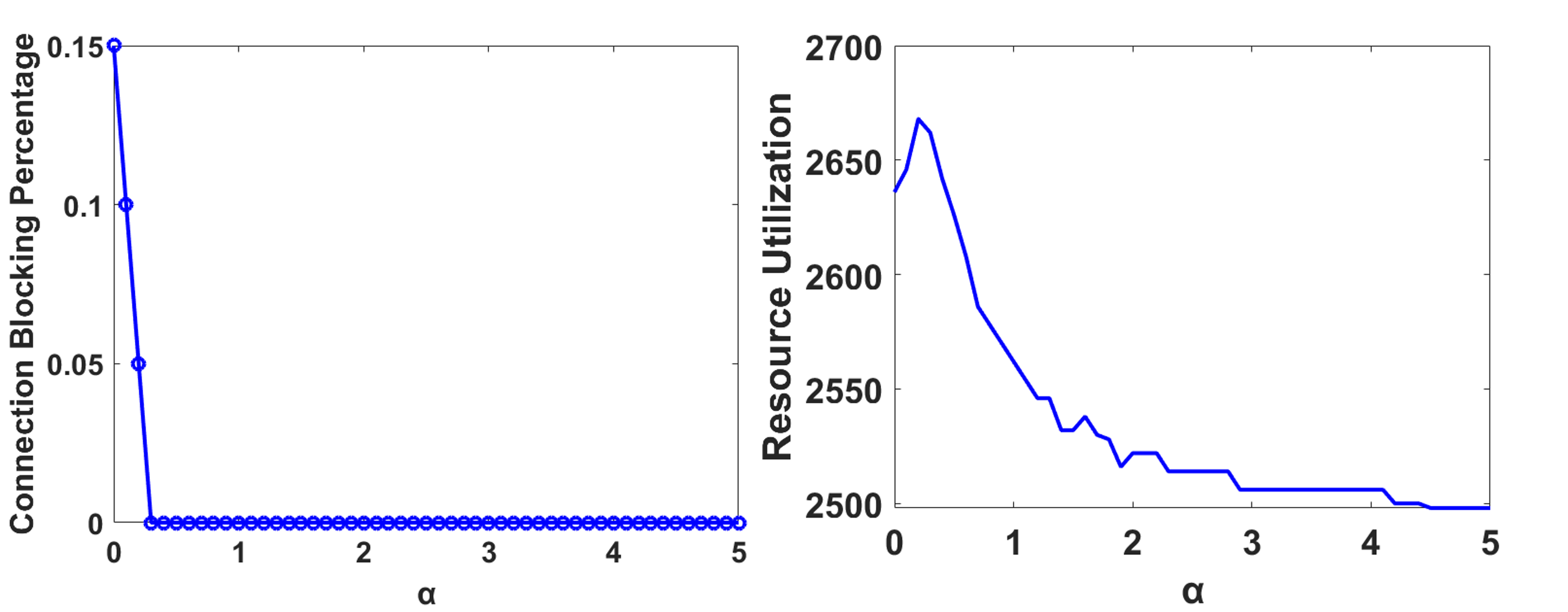}
\caption{(a) Connection blocking percentage vs $\alpha$. (b) Resource utilization in number of FSs vs $\alpha$.}
\label{f0}
\end{center}
\end{figure}

\begin{figure}[h!]
\begin{center}
\includegraphics[width=0.91\columnwidth]{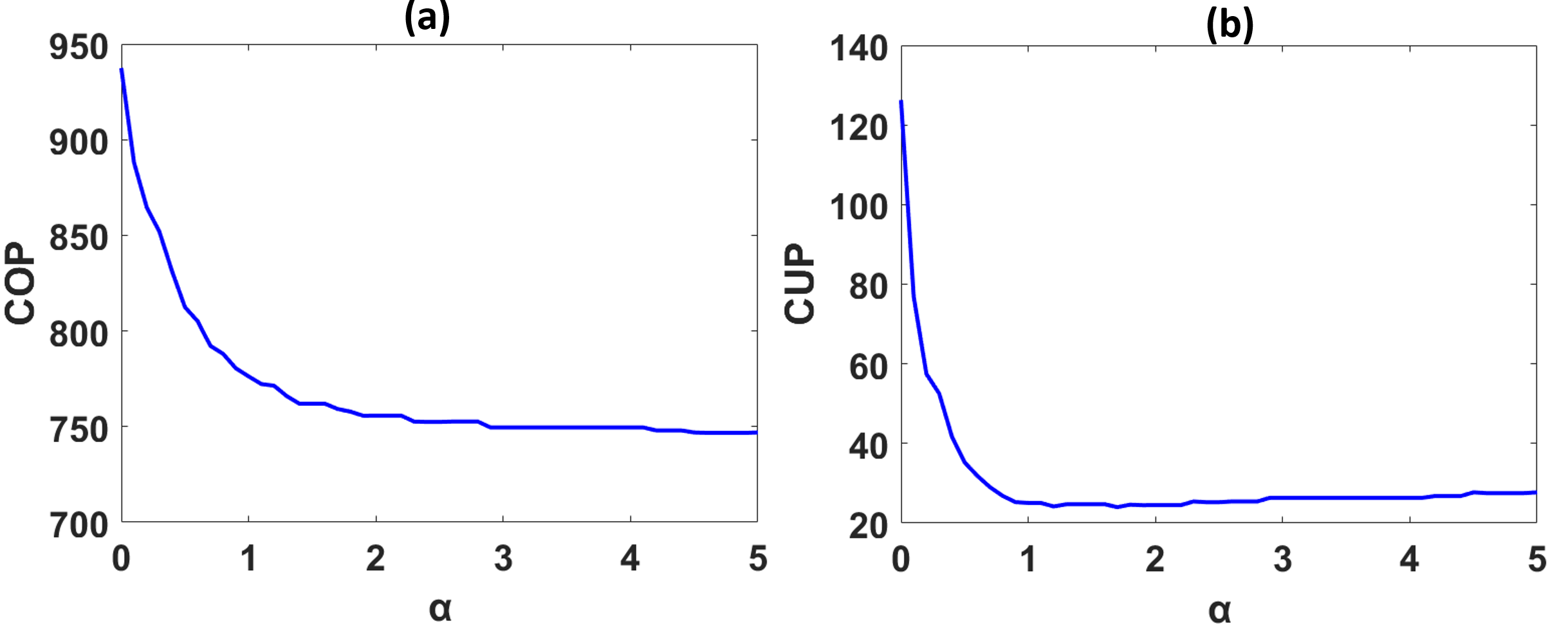}
\caption{(a) $COP(\alpha)$ in number of FSs. (b) $CUP(\alpha)$ in number of FSs.}
\label{f1}
\end{center}
\end{figure}

\begin{figure}[h!]
\begin{center}
\includegraphics[width=0.91\columnwidth]{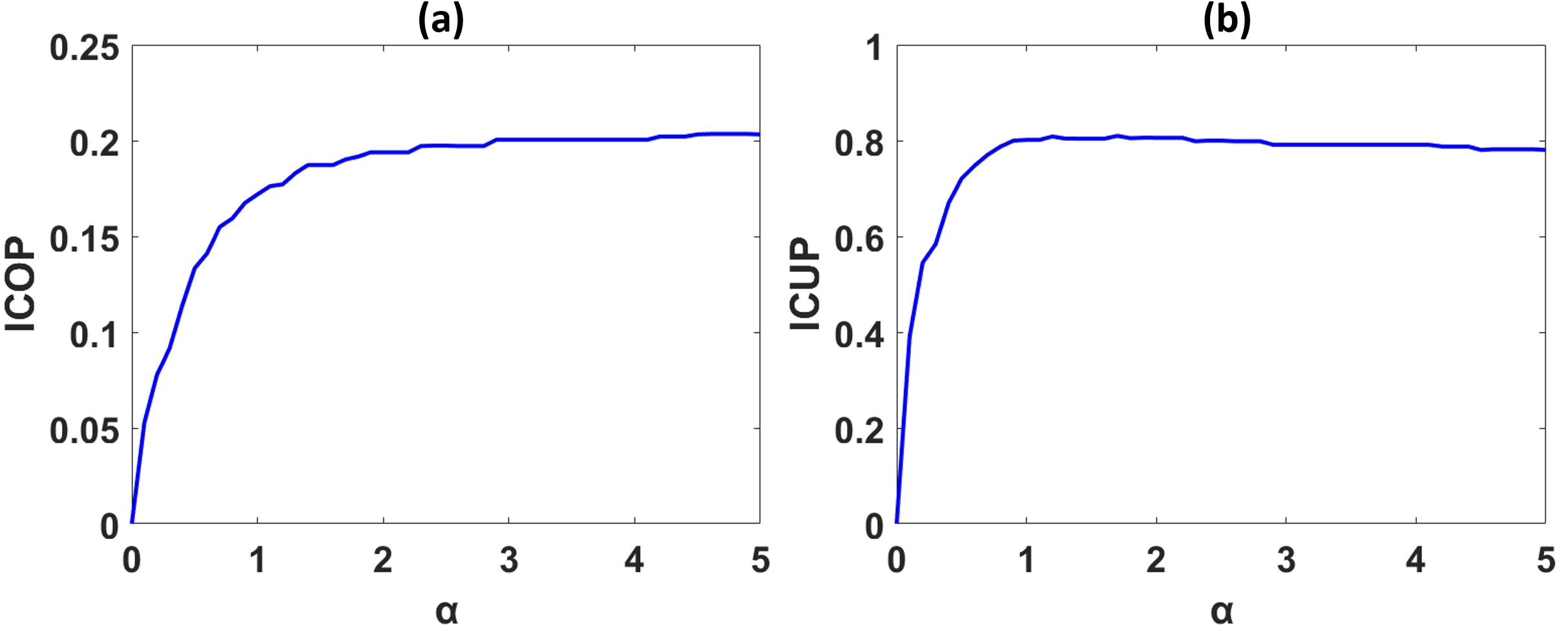}
\caption{(a) $ICOP(\alpha)$. (b) $ICUP(\alpha)$.}
\label{f2}
\end{center}
\end{figure}

\begin{figure}[h!]
\begin{center}
\includegraphics[width=0.91\columnwidth]{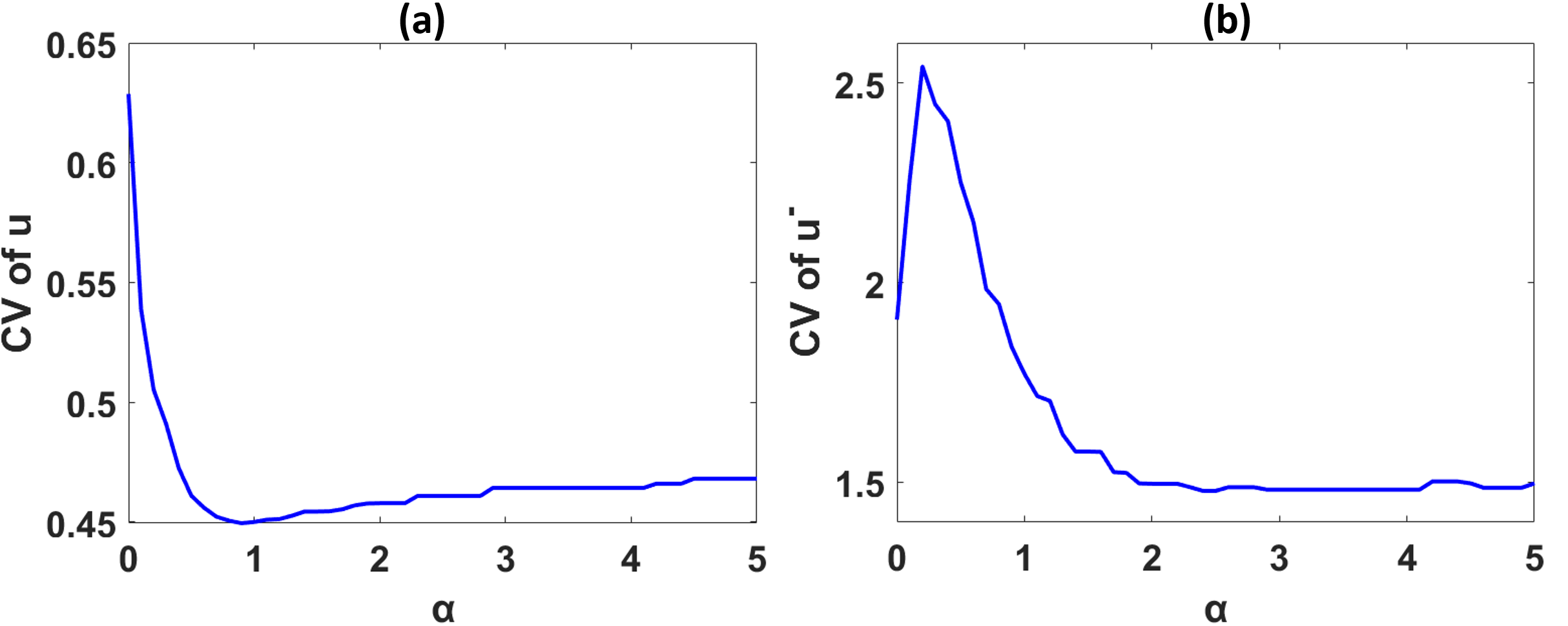}
\caption{(a) CV($\alpha$) of spectrum allocations $u$. (b) CV($\alpha$) of unserved traffic $u^-$.}
\label{f3}
\end{center}
\end{figure}

The $\alpha$-fair RSA ILP is solved in a MATLAB machine with a CPU $@2.60$GHz and $8$GB RAM. Parameter $\alpha$ was examined for the range of values $[0,5]$ with a $0.1$ step, resulting in $50$ $\alpha$-fair allocations $\pi(\alpha)$. In total, $7.5$ hours were required to solve all $50$ ILPs, with the running time of each ranging between $1.5$ hrs and $50$ sec. The resulting $\alpha$-fair allocations were evaluated according to the connection blocking percentage and resource utilization (Fig.~\ref{f0}), COP and CUP (Fig.~\ref{f1}), ICOP and ICUP~(Fig.~\ref{f2}), and CVs of $u$ and $u^-$ (Fig.~\ref{f3}).  

The results indicate that as $\alpha$ increases, connection blocking percentage reduces (Fig.~\ref{f0}(a)). The highest connection blocking is observed for the utilitarian allocation ($\alpha=0$). This is reasonable, as for $\alpha=0$ the objective is to maximize the aggregated network throughput, favoring the connections with higher bit-rate tendencies, and consequently blocking connections with lower bit-rate tendencies. Even though zero connection blocking is achieved for $\alpha=0.3$, results regarding resource utilization, COP/CUP, and CV of $u$ and $u^-$, suggest that $\alpha$ values greater than $0.3$ achieve fairer allocations; an indicator that zero connection blocking alone does not imply fairness.   

In particular, Fig.~\ref{f0}(b) suggests that as $\alpha$ increases, the available spectrum is more efficiently utilized, as the resources required for serving all the connections keep decreasing. Furthermore, as $\alpha$ increases, both COP and CUP tend to decrease (Fig.~\ref{f1}), striking a better balance between the COP and CUP effects. Specifically, for the larger $\alpha$ values examined, COP improves by approximately $20\%$ (Fig.~\ref{f2}(a)) and CUP improves by approximately $80\%$ (Fig.~\ref{f2}(b)). In particular, for our problem setting, approximately the highest improvement in both ICUP and ICOP is achieved around $\alpha=1$, indicating that proportional fairness can be considered as a sufficient solution (if one considers only the ICUP/ICOP measures).       

Furthermore, Fig.~\ref{f3} shows that the CVs of both $u$ and $u^-$ tend to decrease as $\alpha$ increases, indicating that the SAs are fairer with respect to both the utilities $u$ and the unserved traffic $u^-$. It is important to note that the CVs of both unserved traffic and utilities do not change for $\alpha$ values greater than $4.5$ (although not shown in the results $\alpha$ values greater than $5$ were investigated), as an egalitarian SA is reached (i.e., the minimum allocated utility did not further increase). Note that the slight increase in the CV of utilities (Fig.~\ref{f3}(a)) for $\alpha$ values greater than $1$, occurs due to the non-contending connections that  are always allocated their peak-rate requests, consequently causing a variation of utilities that is slightly larger around the mean allocated utilities, as the variability of the utilities of contending connections decreases. Further, the increase in the CV of unserved traffic (Fig.~\ref{f3}(b)) for $\alpha$ values ranging from $0$ to $0.2$, is caused due to the blocked connections in that range (Fig.~\ref{f0}) that differently affect the CV of unserved traffic depending on the traffic demand behavior (i.e., mean and standard deviation) of the exact connections that were blocked.

Importantly, the CVs of both the utilities and the unserved traffic allow the network operator to decide which $\alpha$-fair allocation best fits specific performance requirements. In particular, Fig.~\ref{f3}(b) suggests that $\alpha$ values greater that $2$ best approximate fairer allocations with respect to $u^-$, while Fig.~\ref{f3}(a) suggests that when $\alpha=1$ (i.e., proportional fairness), an allocation that is fairer with respect to $u$ is achieved. Note that proportional fairness also achieves the highest ICUP (Fig.~\ref{f2}(b)). As, however, the network operator may be more interested in approximating an egalitarian SA, achieving at the same time, as similar as possible levels of $u_i^-$, $\alpha$ values greater than $2$ may be preferable. Specifically, $\alpha \geq 2$ achieves zero connection blocking (Fig.~\ref{f0}(a)), only affects resource utilization by approximately $5\%$ (Fig.~\ref{f0}(b)), and achieves approximately $20\%$ ICOP and $80\%$ ICUP (Fig.~\ref{f2}). Furhtermore, CV of $u$ indicates that the fairest SA with respect to the amount of allocated utilities does not imply that the SA is fairest with respect to the connections' QoS. CV of $u^-$, on the other hand, indicates that the fairest QoS-based SA is best approximated by an egalitarian SA (i.e., CV of $u^-$ decreases as $\alpha \rightarrow \infty$).

\section{Conclusion}\label{CON}
In this work, we proposed an $\alpha$-fair RSA framework that is based on the maximization of the social welfare function of the normalized utilities and allows a set of $\alpha$-fair allocations to be computed and evaluated in advance for reconfigurable EONs under modeled tidal traffic. This framework allows for the examination of several existing measures (i.e., connection blocking, resource utilization, CV of utilities), as well as the emergence of new measures (i.e., CV of unserved traffic, COP/CUP, ICOP/ICUP) and their exploitation by a network operator so as to decide on the $\alpha$-fair allocation that best fits the performance criteria of both the connections and the network. Importantly, we show that the new measures and the $\alpha$-fair RSA algorithm achieve to approximate an egalitarian SA that it is also fairer with respect to the QoS for all connections. The egalitarian SA significantly improves connection blocking, expected COP (up to $20\%$), and CUP (up to $80\%$), as opposed to the widely used utilitarian allocation. The development of $\alpha$-fair heuristic approaches for larger problem sizes, as well as techniques for jointly solving the $\alpha$-fair RSA problem while also considering of the various modulation formats are planned as future work.     

\section*{Acknowledgment}
This work has been supported by the European Union's Horizon 2020 research and innovation programme under grant agreement No 739551 (KIOS CoE) and from the Government of the Republic of Cyprus through the Directorate General for European Programmes, Coordination and Development.

\end{document}